\RequirePackage{lineno}
\documentclass[%
    reprint,
    superscriptaddress,
    amsmath,amssymb,
    aps,
    prl,
]{revtex4-1}

\usepackage{graphicx}
\usepackage{dcolumn}
\usepackage{bm}
\usepackage{color}
\usepackage{multirow}
\usepackage{hyperref}
\graphicspath{{./}{./figures/}}

\begin{document}

\title{Characteristics of the diffuse astrophysical electron and tau neutrino flux with six years of IceCube high energy cascade data}

\affiliation{III. Physikalisches Institut, RWTH Aachen University, D-52056 Aachen, Germany} 
\affiliation{Department of Physics, University of Adelaide, Adelaide, 5005, Australia} 
\affiliation{Dept. of Physics and Astronomy, University of Alaska Anchorage, 3211 Providence Dr., Anchorage, AK 99508, USA} 
\affiliation{Dept. of Physics, University of Texas at Arlington, 502 Yates St., Science Hall Rm 108, Box 19059, Arlington, TX 76019, USA} 
\affiliation{CTSPS, Clark-Atlanta University, Atlanta, GA 30314, USA} 
\affiliation{School of Physics and Center for Relativistic Astrophysics, Georgia Institute of Technology, Atlanta, GA 30332, USA} 
\affiliation{Dept. of Physics, Southern University, Baton Rouge, LA 70813, USA} 
\affiliation{Dept. of Physics, University of California, Berkeley, CA 94720, USA} 
\affiliation{Lawrence Berkeley National Laboratory, Berkeley, CA 94720, USA} 
\affiliation{Institut f{\"u}r Physik, Humboldt-Universit{\"a}t zu Berlin, D-12489 Berlin, Germany} 
\affiliation{Fakult{\"a}t f{\"u}r Physik {\&} Astronomie, Ruhr-Universit{\"a}t Bochum, D-44780 Bochum, Germany} 
\affiliation{Universit{\'e} Libre de Bruxelles, Science Faculty CP230, B-1050 Brussels, Belgium} 
\affiliation{Vrije Universiteit Brussel (VUB), Dienst ELEM, B-1050 Brussels, Belgium} 
\affiliation{Dept. of Physics, Massachusetts Institute of Technology, Cambridge, MA 02139, USA} 
\affiliation{Dept. of Physics and Institute for Global Prominent Research, Chiba University, Chiba 263-8522, Japan} 
\affiliation{Dept. of Physics and Astronomy, University of Canterbury, Private Bag 4800, Christchurch, New Zealand} 
\affiliation{Dept. of Physics, University of Maryland, College Park, MD 20742, USA} 
\affiliation{Dept. of Astronomy, Ohio State University, Columbus, OH 43210, USA} 
\affiliation{Dept. of Physics and Center for Cosmology and Astro-Particle Physics, Ohio State University, Columbus, OH 43210, USA} 
\affiliation{Niels Bohr Institute, University of Copenhagen, DK-2100 Copenhagen, Denmark} 
\affiliation{Dept. of Physics, TU Dortmund University, D-44221 Dortmund, Germany} 
\affiliation{Dept. of Physics and Astronomy, Michigan State University, East Lansing, MI 48824, USA} 
\affiliation{Dept. of Physics, University of Alberta, Edmonton, Alberta, Canada T6G 2E1} 
\affiliation{Erlangen Centre for Astroparticle Physics, Friedrich-Alexander-Universit{\"a}t Erlangen-N{\"u}rnberg, D-91058 Erlangen, Germany} 
\affiliation{Physik-department, Technische Universit{\"a}t M{\"u}nchen, D-85748 Garching, Germany} 
\affiliation{D{\'e}partement de physique nucl{\'e}aire et corpusculaire, Universit{\'e} de Gen{\`e}ve, CH-1211 Gen{\`e}ve, Switzerland} 
\affiliation{Dept. of Physics and Astronomy, University of Gent, B-9000 Gent, Belgium} 
\affiliation{Dept. of Physics and Astronomy, University of California, Irvine, CA 92697, USA} 
\affiliation{Karlsruhe Institute of Technology, Institut f{\"u}r Kernphysik, D-76021 Karlsruhe, Germany} 
\affiliation{Dept. of Physics and Astronomy, University of Kansas, Lawrence, KS 66045, USA} 
\affiliation{SNOLAB, 1039 Regional Road 24, Creighton Mine 9, Lively, ON, Canada P3Y 1N2} 
\affiliation{Department of Physics and Astronomy, UCLA, Los Angeles, CA 90095, USA} 
\affiliation{Department of Physics, Mercer University, Macon, GA 31207-0001, USA} 
\affiliation{Dept. of Astronomy, University of Wisconsin, Madison, WI 53706, USA} 
\affiliation{Dept. of Physics and Wisconsin IceCube Particle Astrophysics Center, University of Wisconsin, Madison, WI 53706, USA} 
\affiliation{Institute of Physics, University of Mainz, Staudinger Weg 7, D-55099 Mainz, Germany} 
\affiliation{Department of Physics, Marquette University, Milwaukee, WI, 53201, USA} 
\affiliation{Institut f{\"u}r Kernphysik, Westf{\"a}lische Wilhelms-Universit{\"a}t M{\"u}nster, D-48149 M{\"u}nster, Germany} 
\affiliation{Bartol Research Institute and Dept. of Physics and Astronomy, University of Delaware, Newark, DE 19716, USA} 
\affiliation{Dept. of Physics, Yale University, New Haven, CT 06520, USA} \affiliation{Dept. of Physics, University of Oxford, Parks Road, Oxford OX1 3PU, UK} 
\affiliation{Dept. of Physics, Drexel University, 3141 Chestnut Street, Philadelphia, PA 19104, USA} 
\affiliation{Physics Department, South Dakota School of Mines and Technology, Rapid City, SD 57701, USA} 
\affiliation{Dept. of Physics, University of Wisconsin, River Falls, WI 54022, USA} 
\affiliation{Dept. of Physics and Astronomy, University of Rochester, Rochester, NY 14627, USA} 
\affiliation{Oskar Klein Centre and Dept. of Physics, Stockholm University, SE-10691 Stockholm, Sweden} 
\affiliation{Dept. of Physics and Astronomy, Stony Brook University, Stony Brook, NY 11794-3800, USA} 
\affiliation{Dept. of Physics, Sungkyunkwan University, Suwon 16419, Korea} 
\affiliation{Institute of Basic Science, Sungkyunkwan University, Suwon 16419, Korea} 
\affiliation{Dept. of Physics and Astronomy, University of Alabama, Tuscaloosa, AL 35487, USA} 
\affiliation{Dept. of Astronomy and Astrophysics, Pennsylvania State University, University Park, PA 16802, USA} 
\affiliation{Dept. of Physics, Pennsylvania State University, University Park, PA 16802, USA} 
\affiliation{Dept. of Physics and Astronomy, Uppsala University, Box 516, S-75120 Uppsala, Sweden} 
\affiliation{Dept. of Physics, University of Wuppertal, D-42119 Wuppertal, Germany} 
\affiliation{DESY, D-15738 Zeuthen, Germany}

\author{M. G. Aartsen}\affiliation{Dept. of Physics and Astronomy, University of Canterbury, Private Bag 4800, Christchurch, New Zealand} \author{M. Ackermann} \affiliation{DESY, D-15738 Zeuthen, Germany} \author{J. Adams} 
\affiliation{Dept. of Physics and Astronomy, University of Canterbury, Private Bag 4800, Christchurch, New Zealand} \author{J. A. Aguilar} \affiliation{Universit{\'e} Libre de Bruxelles, Science Faculty CP230, B-1050 Brussels, Belgium} \author{M. Ahlers} \affiliation{Niels Bohr Institute, University of Copenhagen, DK-2100 Copenhagen, Denmark} \author{M. Ahrens} \affiliation{Oskar Klein Centre and Dept. of Physics, Stockholm University, SE-10691 Stockholm, Sweden} \author{C. Alispach} \affiliation{D{\'e}partement de physique nucl{\'e}aire et corpusculaire, Universit{\'e} de Gen{\`e}ve, CH-1211 Gen{\`e}ve, Switzerland} \author{K. Andeen} \affiliation{Department of Physics, Marquette University, Milwaukee, WI, 53201, USA} \author{T. Anderson} \affiliation{Dept. of Physics, Pennsylvania State University, University Park, PA 16802, USA} \author{I. Ansseau} \affiliation{Universit{\'e} Libre de Bruxelles, Science Faculty CP230, B-1050 Brussels, Belgium}
% ...

\collaboration{IceCube Collaboration}
\email{analysis@icecube.wisc.edu}

\date{\today}

\begin{abstract}
We report on the first measurement of the astrophysical neutrino flux using particle showers (cascades) in IceCube data from 2010 -- 2015. 
Assuming standard oscillations, the astrophysical neutrinos in this dedicated cascade sample are dominated ($\sim 90 \%$) by electron and tau flavors.
The flux, observed in the sensitive energy range from $16\,\textnormal{TeV} $ to $2.6\,\textnormal{PeV}$, is consistent with a single power-law  model 
as expected from Fermi-type acceleration of high energy particles at astrophysical sources. 
We find the flux spectral index to be $\gamma=2.53\pm0.07$  and a flux normalization for each neutrino flavor 
of  $\phi_{astro} = 1.66^{+0.25}_{-0.27}$ at $E_{0} = 100\, \textnormal{TeV}$,
in agreement with IceCube's complementary muon neutrino results and with all-neutrino flavor fit results. 
In the measured energy range we reject spectral indices $\gamma\leq2.28$ at $\ge3\sigma$ significance level.
Due to high neutrino energy resolution and low atmospheric neutrino backgrounds,
this analysis provides the most detailed characterization of the neutrino flux at energies below 
$\sim100\,{\rm{TeV}}$ compared to previous IceCube results.
Results from fits assuming more complex neutrino flux models suggest a flux softening at high energies and a flux hardening at low energies (p-value  $\ge 0.06$).
The sizable and smooth flux measured  below $\sim 100\,{\rm{TeV}}$  remains a puzzle.
In order to not violate the isotropic diffuse gamma-ray
background as measured by the Fermi-LAT, it suggests  the existence of astrophysical neutrino sources characterized by dense environments which are opaque to gamma-rays.
\end{abstract}


\maketitle

\noindent
In 2013 IceCube discovered a diffuse and isotropic flux of neutrinos of astrophysical  origin~\cite{ehe-prl-2013,HESE2y,HESE3y}.   
In 2018, an Active Galactic Nucleus (AGN) with a relativistic jet pointing towards the Earth, the blazar TXS 0506+056,
was identified as the first possible extra-galactic source of astrophysical neutrinos and  cosmic ray accelerator
~\cite{MMessenger-170922A,TXS-AlertPrior}.  
In diffuse neutrino flux measurements one aims to  gain insights  
into astrophysical neutrino production mechanisms,  typically associated with cosmic ray acceleration at the source, and 
interactions either with surrounding gas  ($pp$) or photons  ($p\gamma$).  
The Fermi shock acceleration mechanism of high energy cosmic rays, in sources such as AGN~\cite{Protheroe,KazEll,Begelman,Stecketal,StecketalE,MannBier},   
predicts the flux of neutrinos to follow a single power law $E^{-\gamma}$ with a baseline spectral index of $\gamma\sim 2$ for strong shocks~\cite{Bell-1978,Gaisser1990}. 
The spectral index and flux normalization factors carry information about neutrino sources and the environment~\cite{Winter2013,murase}.    
Different production mechanisms together with energy losses of pions and muons lead, depending on energy, 
to different neutrino flavor compositions at sources and,  after neutrino oscillations over astrophysical distances, at the Earth~\cite{LearnedPakvasa,Athar2006,KashtiWaxman,Klein2013,Lipari2007,Bustamante2015,Esmaili2009,nufit,nufit2}.  
The main goal of  astrophysical neutrino flux measurements is a characterization of its energy dependence in a flavor dependent way 
and in a wide energy range~\cite{Tracks6y,Tracks10y,StartingTracks5y,HESE75y,MESE2y,global-fit,Antares-diffuse,gvd}, 
relevant for ultra high energy cosmic rays and QCD physics.  
Since the diffuse Galactic emission component, based on models of galactic particle propagation and interactions~\cite{kra},    
is sub-dominant~\cite{Antares-IceCube-Galactic,PS-Cascades7yr}, 
of particular interest is the energy range $\sim10-100$ TeV.  
In this energy region, hardly accessible to muon neutrinos, 
several source models, including AGN cores~\cite{agn-cores1,agn-cores2} 
predict a sizable energy dependent flux.  
In this paper we present the first results on the astrophysical flux of electron and tau neutrinos determined 
with $6$ years of IceCube data.

IceCube is a neutrino observatory comprising $5160$  Digital Optical Modules (DOMs)~\cite{doms} distributed over one cubic kilometer  in the Antarctic ice. 
Charged particles, which are produced in neutrino interactions, emit Cherenkov light while propagating through the ice.  
The Cherenkov light detected by the optical sensors forms three types of patterns, 
muon tracks (starting inside or going through the detector) and cascades.  Single cascades are electromagnetic  and/or hadronic particle showers produced by 
(i) electron or  low energy tau neutrinos scattering inelastically off target nucleons through a $W$ boson (ii) neutrinos of all flavors scattering inelastically off target nucleons  through a $Z$ boson  or 
(iii) electron anti-neutrinos interacting with atomic electrons to form a  $W^-$ boson,  the Glashow Resonance~\cite{gr}. 
Although the angular resolution of cascades is limited 
($> 8^{\circ}$)~\cite{PS-Cascades7yr},  
their energy resolution ($\sim15\%$)~\cite{EnergyReco2014} as well as  
their low atmospheric neutrino background make the cascade channel particularly well suited for measuring and characterizing 
the energy dependent astrophysical neutrino flux~\cite{beacom2004}.

We analyzed $6$ years of IceCube cascade data, collected in $2010-2015$. 
We used IceCube Monte Carlo simulation packages to simulate the cosmic ray background with CORSIKA~\cite{corsika} and single muons from cosmic rays with MuonGun~\cite{JakobThesis}. 
For the cosmic ray primary flux  
we used the Gaisser-H3a~\cite{Gaisser-H3a} model    
and SIBYLL 2.1~\cite{SIBYLL21} as the hadronic interaction model. 
High energy neutrino interactions were generated with the NuGen software package 
based on ~\cite{anis}.  
The total $\nu$N deep inelastic scattering cross section is from~\cite{sarkar}. 
Astrophysical neutrino event selection efficiencies were tested assuming as baseline an $E^{-2}$ flux with equal numbers of neutrinos and anti-neutrinos, and 
with an equal neutrino flavor mixture at Earth: $(\nu_e : \nu_{\mu} : \nu_{\tau})_E = (\bar{\nu}_e : \bar{\nu}_{\mu} : \bar{\nu}_{\tau} )_E = 0.5 : 0.5 : 0.5$.
The conventional atmospheric neutrino  flux from pion and kaon decays was modeled according to~\cite{HKKMS06}, with primary cosmic ray flux modifications according to the Gaisser-H3a model~\cite{Gaisser-H3a}.  
It is in agreement, in the energy range relevant to this analysis  $E > 400\,{\mathrm{GeV}}$, with the atmospheric neutrino flux measurements by Super-Kamiokande~\cite{SuperK2016},  AMANDA-II~\cite{AMANDA2009,AMANDA2010},  IceCube~\cite{ICAtm2011, ICAtm2013,ICAtm2015},  and ANTARES~\cite{ANTARES2013}.  
Atmospheric neutrinos originating from the decays of charm or heavier mesons produced in air-showers,
so-called prompt neutrinos, are yet to be detected.
We used the BERSS model~\cite{BERSS}
to predict the contribution from prompt neutrinos to the total neutrino flux,   
and the atmospheric neutrino self veto effect calculations from~\cite{veto2}, tuned  to match our full CORSIKA Monte Carlo simulations.

\begin{figure}[t!]
\centering
\includegraphics[width=0.49\textwidth]{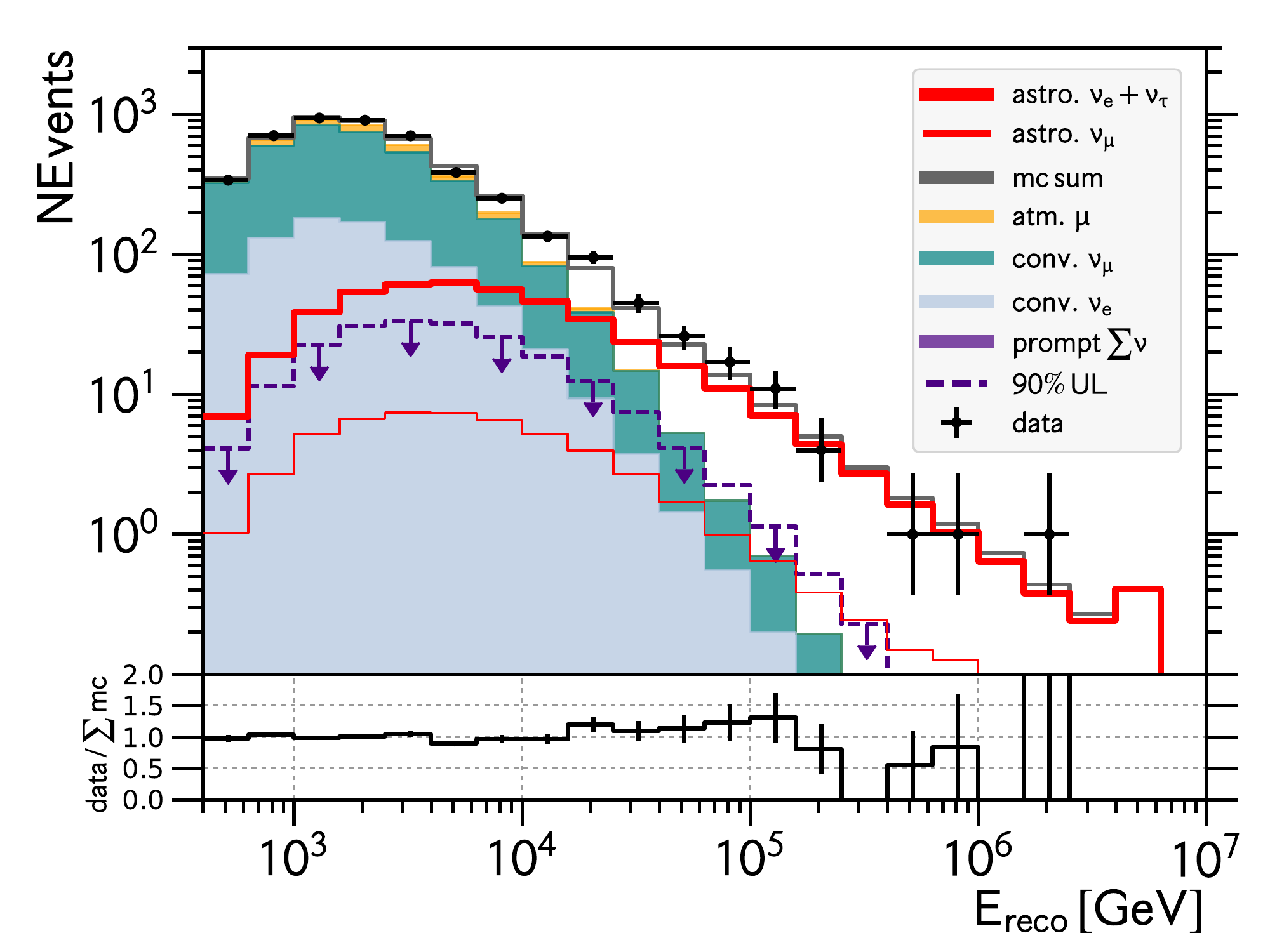} \\
\includegraphics[width=0.49\textwidth]{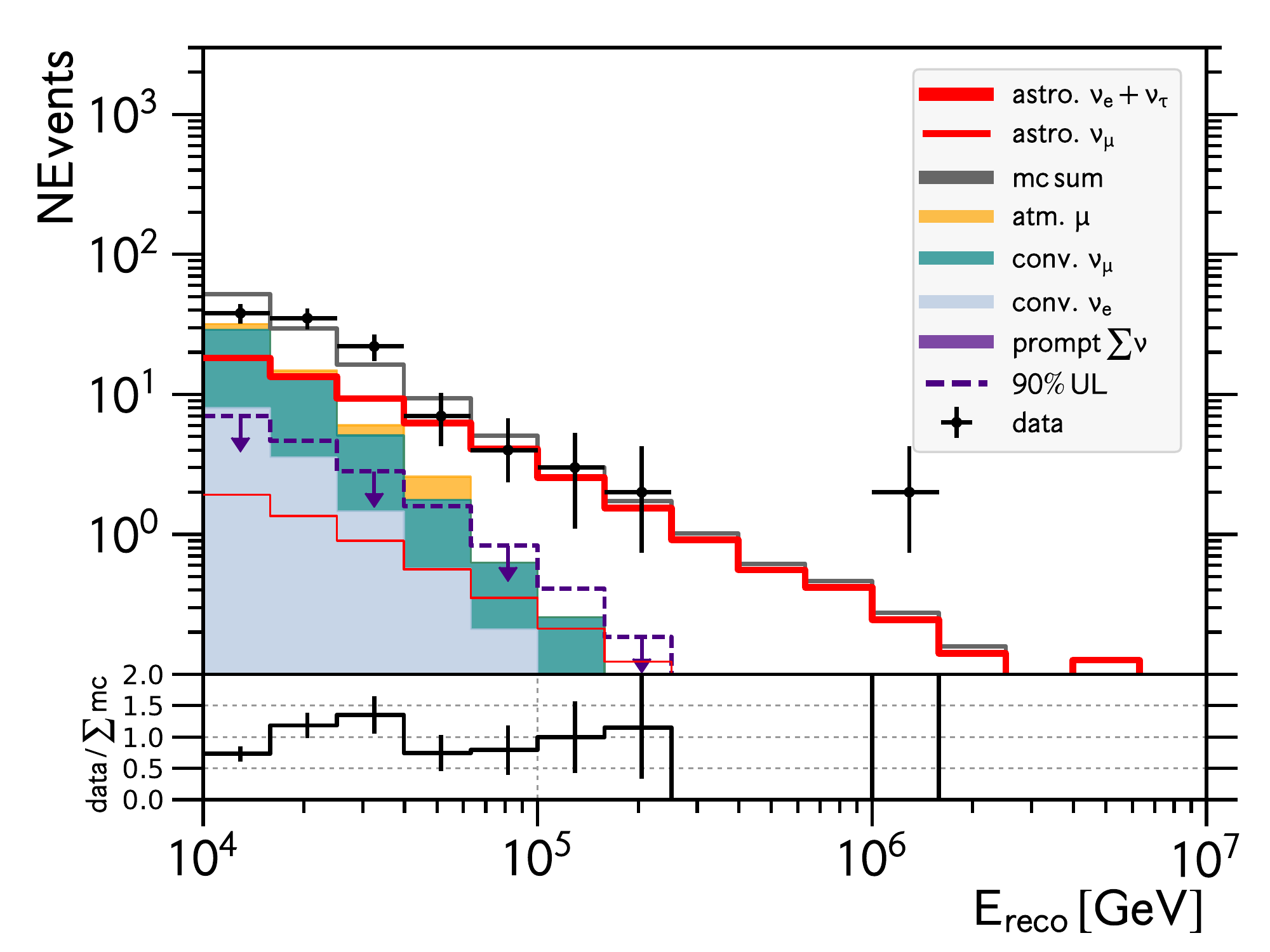}
\caption{
Reconstructed cascade energy distribution. Black points are  data, with statistical uncertainties, acquired during
the observation period.  Continuous lines are  Monte Carlo simulations as labeled in the legend.   
The atmospheric background histograms are stacked (filled colors).
Shown are best fit distributions assuming single power-law model of the astrophysical neutrino flux (Tab.~\ref{tab3}). 
Top: data from $2012-2015$ ({\em{Sample-B}}). Bottom: data from $2010-2011$ ({\em{Sample-A}}).
}
\label{fig1}
\end{figure}

The analyzed data consists of two sets:  $2010-2011$ (2 years, {\em{Sample-A}})~\cite{Cascades2y} and $2012-2015$ (4 years, {\em{Sample-B}})~\cite{Cascades4y,HansThesis,YiqianThesis}.  
Events from both samples passed IceCube's  dedicated online {\em{cascade filter}},  which utilizes results of simple muon and cascade reconstruction algorithms.     
The cascade filter reduces the cosmic ray background rate from $\sim2.7\,{\mathrm{kHz}}$  to $\sim\,30{\mathrm{Hz}}$,  while retaining 
$\sim\,90\%$ of the expected astrophysical neutrinos and $\sim\,70\%$  of the conventional atmospheric neutrinos.  
In order to further reduce backgrounds and ensure high neutrino induced cascade signal efficiencies and good cascade energy resolution,  a fiducial volume selection on the reconstructed 
cascade vertex position was imposed.  
A straight cut selection method was used to select signal cascades  in {\em{Sample-A}}  ($E\,>\,10\,{\mathrm{TeV}}$)~\cite{Cascades2y}
and in the high energy ($E\,>\,60\,{\mathrm{TeV}}$) subset of {\em{Sample-B}}~\cite{Cascades4y,YiqianThesis}.   
It builds on methods developed 
in previous IceCube searches  dedicated to astrophysical cascades performed with partial detector configurations during IceCube construction periods~\cite{CascadesIC22,CascadesIC40,CascadesIC59}.  
A significant improvement was achieved by applying 
 a Boosted Decision Tree~\cite{bdt}  method  in the low energy ($\sim\,400\,{\mathrm{GeV}} < E < 60\,{\mathrm{TeV}}$) subset of  {\em{Sample-B}}
to  classify events according to their topology into muon track background, signal neutrino induced cascades and muon starting track events~\cite{Cascades4y,HansThesis}.  
The obtained cascade sample has  low ($8\%$) muon background contamination.   Lowering the energy threshold from $10\,{\mathrm{TeV}}$ ({\em{Sample A}}) to  $\sim\,400\,{\mathrm{GeV}}$  ({\em{Sample B}}) substantially reduces  systematic uncertainties in this measurement. Reconstructed cascade energy distributions for {\em{Sample-A}} and for {\em{Sample-B}}  after  all selections are shown as black points in Fig.~\ref{fig1}.   
About $60\%$ of the cascades identified in this analysis and with reconstructed energies above 60 TeV do not contribute to the High Energy Starting Events (HESE)~\cite{HESE75y} cascade data sample for the same period ($2010-2015$).
Monte Carlo simulations show  that at $10$ TeV this analysis increases the total expected  number of electron neutrinos by a factor of  $\sim10$ compared to the Medium Energy Starting Events (MESE) analysis~\cite{MESE2y}.

We determined the astrophysical neutrino flux, characterized by parameters $\boldsymbol{\theta_r}$, 
by maximizing a binned poisson likelihood  $L\left(\boldsymbol{\theta_r},\,\boldsymbol{\theta_s}\,|\,\mathbf{n}\right) $. The  $\boldsymbol{\theta_s}$  are the nuisance parameters, 
and $\mathbf{n} = (n_1, ...., n_m)$ is the vector of observed event counts $n_i$ in the $i^{th}$ bin. 
The fit  was performed in bins of three observables:  event type (cascade, muon track, muon starting track),  reconstructed  energy,  and reconstructed zenith angle in the range $0 - \pi$,  as shown in Table~\ref{tab:binning}.  
In this analysis, the log-likelihood function is defined, up to a constant,  as:  

\begin{small} 
\vspace*{-0.62cm}
\begin{align*}
	&\log L\left(\boldsymbol{\theta_r},\,\boldsymbol{\theta_s}\,|\,\mathbf{n}\right)  =  \sum_{i=1}^{m}\left[ n_i\log\mu_i\left(\boldsymbol{\theta_r,\theta_s}\right)-\mu_i\left(\boldsymbol{\theta_r,\theta_s}\right)\right] +  \\ \newline
	& \frac{1}{2}\left[\hspace*{-0.1cm}\left(\frac{\epsilon_{eff}^{DOM}-\hat{\epsilon}_{eff}^{DOM} }{\sigma_\epsilon^{DOM}}\right)^{\hspace*{-0.1cm}2}\hspace*{-0.1cm}+\hspace*{-0.1cm}\left(\frac{\epsilon^{HI}_{abs}-\hat{\epsilon}^{HI}_{abs}}{\sigma_\epsilon^{HI}}\right)^{\hspace*{-0.1cm}2}\hspace*{-0.1cm}+\hspace*{-0.1cm}\left(\frac{\Delta\gamma_{CR}-\hat{\Delta}\gamma_{CR}}{\sigma_{\Delta\gamma_{CR}}}\right)^{\hspace*{-0.1cm}2}\right]  
 \end{align*} 
\vspace*{-0.25cm}
 \begin{align}
	 & + \frac{1}{2}\left(\boldsymbol{\epsilon^{BI}}-\boldsymbol{\hat{\epsilon}^{BI}}\right)^T\boldsymbol{\Sigma}_{BI}^{-1}\left(\boldsymbol{\epsilon^{BI}}-\boldsymbol{\hat{\epsilon}^{BI}} \right). 
\label{eqL}
\end{align}
\end{small}
\begin{table}[t!]
\small
\centering
\begin{tabular}{lccccc}\hline
     Sample \&  & Energy &  Energy   & Zenith & Zenith  \\  \\[-2.5ex] 
            Event Type          &   NBins        & Range   &  NBins & Range \\\hline \hline \\[-2.5ex] 
     A  \textbf{cascade}& $15$ &$4.0\,-\,7.0$ & $3$ & $0-\pi$  \\
     B  \textbf{cascade}& $22$ &$2.6\,-\,7.0$ & $3$ & $0-\pi$ \\ 
     B  $\mu$ starting track& $11$ &$2.6\,-\,4.8$ & $1$ &  $0-\pi$ \\
     B  $\mu$  track& $1$ &  $2.6\,-\,4.8$ & $1$ & $0-\pi$  \\\hline
\end{tabular}
\caption{The binning of observables (reconstructed energy and zenith) used in the maximum likelihood fit. Energy ranges are given in logarithmic units, $\log_{10}E/\mathrm{GeV}$, and zenith ranges are given in radians.  The three bins' ranges  in $\cos({\rm{Zenith}})$ are $(-1, 0.2, 0.6, 1)$ }
\label{tab:binning}
\end{table}
\noindent
The expected, from Monte Carlo simulations, number of events in the $i^{th}$  bin is  defined as $\mu_i =  \mu_i^{atm.\mu} + \mu_i^{conv.\nu} + \mu_i^{prompt\nu} + \mu_i^{astro.\nu}$,
the sum of background cosmic ray muons, conventional and prompt atmospheric neutrinos, and  astrophysical neutrinos. 
The nuisance parameters $\boldsymbol{\theta_s}$  contribute additive penalty terms to the log-likelihood function, Eq.~(\ref{eqL}). They  
account for detector related systematic uncertainties,   comprised of 
the DOM optical efficiency, $\epsilon_{eff}^{DOM}$, 
optical properties (scattering and absorption length) of the bulk ice (BI), $\epsilon^{BI}_{scat}$ and $\epsilon^{BI}_{abs}$,  and of the re-frozen drilled hole ice (HI), $\epsilon^{HI}_{scat}$. 
The bivariate covariance matrix $\boldsymbol\Sigma_{BI} $ takes into account correlations between the two components of $\boldsymbol\epsilon^{BI}= (\epsilon^{BI}_{scat}$, $\epsilon^{BI}_{abs}$). 
Other systematic uncertainties are due to uncertainties on the cosmic ray flux index $\Delta\gamma_{CR}$,  on the flux normalizations of the cosmic ray muon $\phi_{muon}$,  atmospheric conventional $\phi_{conv}$ and prompt 
$\phi_{prompt}$ neutrino backgrounds.  Uncertainties in the atmospheric neutrino flux prediction related to hadronic interaction models~\cite{SIBYLL23c,EPOS-LHC,QGSJETIIv04,DPMJETIII171} 
have been studied  using the MCEq~\cite{mceq} package.  They were found small and thus neglected.    

\begin{figure}[t!]
\centering
\includegraphics[width=0.44\textwidth]{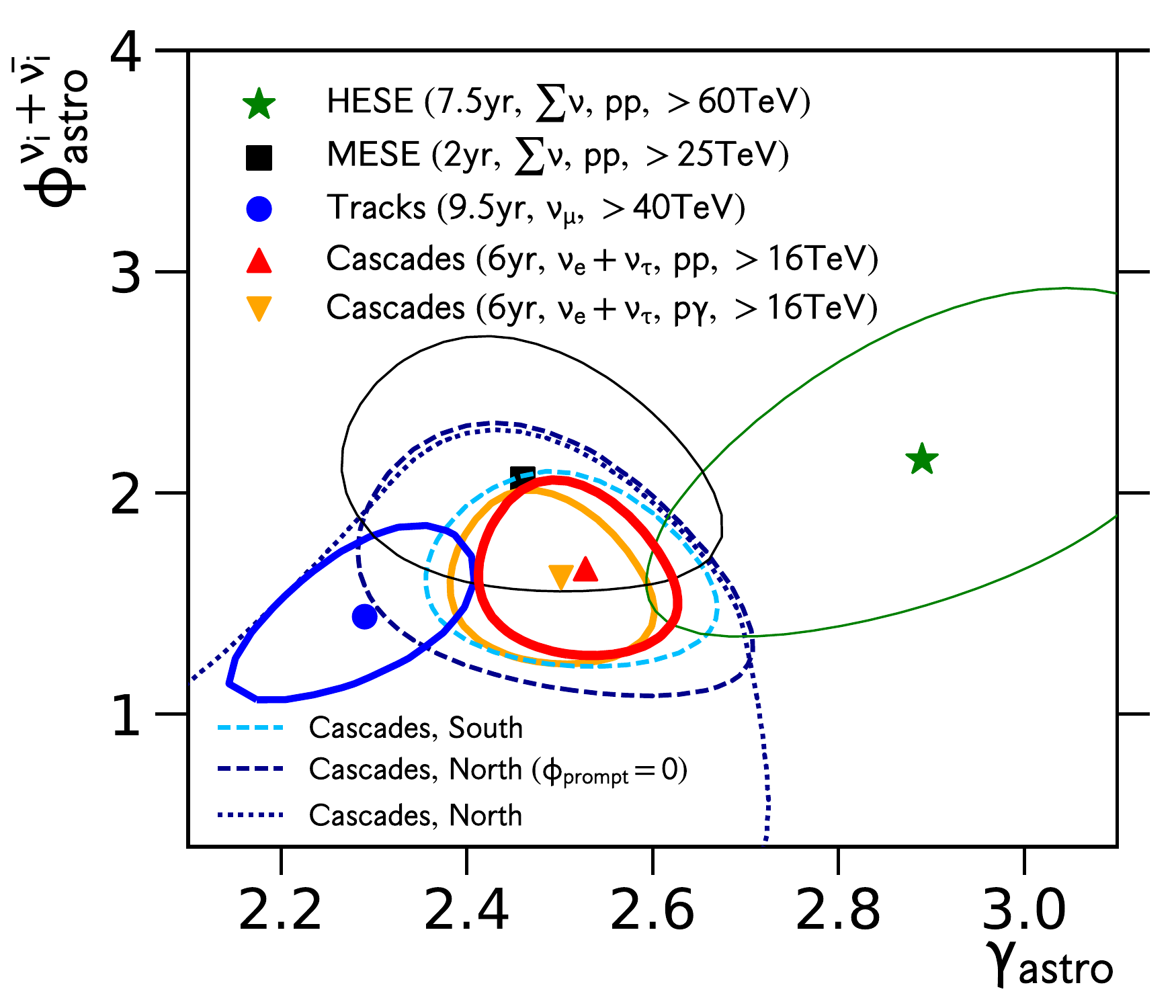} \\
\caption{$68 \%$ C.L. profile likelihood contours for the single power-law astrophysical neutrino flux fit parameters, the flux normalization (per neutrino flavor) and the spectral index. Shown are results for the combined 2010-2015  (6 years) cascade analysis. Red (yellow) curves are obtained assuming $pp$ ($p\gamma$) neutrino production mechanism at the source, respectively.
Other IceCube results are shown as blue, green and gray curves for $\nu_{\mu}$~\cite{Tracks10y} and for all-neutrino flavor   
HESE~\cite{HESE75y} and MESE~\cite{MESE2y} analyses. 
}\label{fig2}
\end{figure}

We performed several fits considering different functional forms of the astrophysical neutrino flux.  
All models assume equal numbers of neutrinos and anti-neutrinos and equal neutrino flavors at Earth.
First we describe the results obtained for the 
single power law flux model: 
\begin{equation}
\Phi_{\textnormal{astro}}^{\nu+{\bar{\nu}}}  \left(E\right) / C_0 =  \phi_{astro} \times \left(E /E_0\right)^{-\gamma},
\end{equation}
\noindent
where $C_0 = 3\times 10^{-18}\,\textnormal{GeV}^{-1}\!\cdot\!\textnormal{cm}^{-2}\!
\cdot\!\textnormal{s}^{-1}\!\cdot\!\textnormal{sr}^{-1}$ and $E_{0} = 100\, \textnormal{TeV}$. 
We find  the following best fit parameters: the flux spectral index $\gamma=2.53\pm0.07$ and  
the  flux normalization for each neutrino flavor
$\phi_{astro} = 1.66^{+0.25}_{-0.27}$ at $E_{0} = 100\, \textnormal{TeV}$. 
The result  for the measured electron and tau neutrino flux  $\Phi_{\textnormal{astro}}^{\nu_e+{\bar{\nu}_e}} +  \Phi_{\textnormal{astro}}^{\nu_{\tau}+{\bar{\nu}_{\tau}}}$  
changes insignificantly, if we include variations in the injected
flavor ratio at astrophysical sources $(\nu_e:\nu_\mu:\nu_\tau)_S=(1-f_\mu^S\,:\,f_\mu^S\,:\,0)$  through an additional nuisance parameter 
$0\leq f_\mu^S \leq1$, 
as shown in the Supplemental Material  Fig. 1 (right)~\cite{Supplement}. The sensitive energy range,
defined as the smallest range where a non-zero astrophysical flux is consistent with the data at $90\%$ C.L.~\cite{HansThesis}, 
ranges from $16\,\textnormal{TeV}$ to $2.6\,\textnormal{PeV}$.  The best fit values of all physics and nuisance fit parameters and their uncertainties are given in Table~\ref{tab3}. 
Figure~\ref{fig1} shows the reconstructed cascade energy distributions for data and for  Monte Carlo simulations with the signal and background contributions scaled according to the best fit values of all fit parameters. The agreement between data and simulations is very good with a goodness-of-fit~\cite{gof} p-value of $0.88$~\cite{HansThesis}.  The number of neutrino events based on the best fit results are shown in Table~\ref{tab4}.
The contribution from astrophysical electron and tau neutrinos to the cascade samples strongly dominates over the small ($12\%$) contribution from astrophysical muon neutrinos.
The  energy and zenith angle dependence of the measured flux  is consistent with expectations for a flux of neutrinos of astrophysical origin. 
The $68\%$ C.L. profile likelihood contours for the correlated spectral index and flux normalization are shown in Fig.~\ref{fig2} as a red curve. 
Similar results (yellow curve, $\gamma=2.50\pm 0.07$  and $\phi_{astro} =1.62^{+0.25}_{-0.27}$ ) 
were obtained under the assumption that the astrophysical neutrino flux originated from the 
$p\gamma$-type source where we used the at-earth flavor ratios,  $(\nu_e : \nu_{\mu} : \nu_{\tau})_E = 0.78: 0.61: 0.61$ and $(\bar{\nu}_e : \bar{\nu}_{\mu} : \bar{\nu}_{\tau} )_E = 0.22 : 0.39 : 0.39$~\cite{nu-flavor-ratios}, and assumed the single power law flux.
No significant difference has been observed between the fluxes from the Northern and Southern skies (dashed cyan and blue lines in Fig.~\ref{fig2}). 
Since the atmospheric self veto effect~\cite{JakobThesis,veto1,veto2,veto3} reduces atmospheric neutrino background in the Southern sky, the astrophysical flux is measured more precisely in the Southern than in the Northern hemisphere,  $\gamma_S=2.52^{+0.10}_{-0.11}$ and  $\gamma_N=2.45^{+0.17}_{-0.36}$ (Tab.~\ref{tab:fits}, hypothesis F). 
\begin{table}[t!]
\small
\centering
\begin{tabular}{lll}\hline
      Parameter  & Prior & Result $\pm 1\sigma$ \\
                 & constraint  & ($<90\%$ upper limit)  \\\hline\hline\\[-2.5ex]
	$\mathbf{\gamma}$ & - & $\mathbf{2.53\pm0.07}$ \\
        $\mathbf{{\phi}_{astro}}$ & - & $\mathbf{1.66^{+0.25}_{-0.27}}$ \\
        $\phi_{conv}$ & - & $(1.07^{+0.13}_{-0.12})\times \Phi_{{\small{\text{HKKMS06}}}}$\\
        $\phi_{prompt}$ & - & $<5.0\times\Phi_{\small{\text{BERSS}}}$ \\
        $\phi_{muon}$ & - & $1.45\pm0.04$ \\
        $\Delta\gamma_{CR}$ & $0.00\pm0.05$ & $0.02\pm0.03$\\
        $\epsilon_{scat}^{BI}$ & $1.00\pm0.07$ & $1.02\pm0.03$ \\
        $\epsilon_{abs}^{BI}$ & $1.00\pm0.07$ & $1.03^{+0.05}_{-0.04}$\\
        $\epsilon_{scat}^{HI}$ & - & $1.72\pm 0.19$\\
       $\epsilon_{eff}^{DOM}$ & $0.99\pm0.10$ & $1.03^{+0.08}_{-0.07}$ \\\hline \hline
\end{tabular}
\caption{
Best fit values and uncertainties for all parameters included in the single power law  fit. 
}
\label{tab3}
\end{table}
\begin{table}[h!]
\small
\centering
\begin{tabular}{lccc}\hline
    Number of Events  & $\nu_{e}$+$\bar{\nu}_{e}$ & $\nu_{\mu}$+$\bar{\nu}_{\mu}$ & $\nu_{\tau}$+$\bar{\nu}_{\tau}$  \\\hline\hline\\[-2.3ex]
     astro.  &  $303^{+46}_{-45}$ & $59^{+8}_{-7}$ & $204^{+28}_{-27}$  \\ [+0.5ex] 
                 & ($127^{+12}_{-12}$)  & ($22^{+2}_{-2}$)  & ($80^{+7}_{-7}$) \\ [+0.5ex] \hline \\[-2.3ex]  
     astro. GR   &  $0.73^{+0.31}_{-0.22}$  & - & - \\ [+0.5ex]  \hline
     atmo. conv. & $851^{+23}_{-23}$  & $2901^{+64}_{-65}$ & -  \\ [1.0ex]
     & ($50^{+3}_{-3}$)  & ($143^{+8}_{-8}$)  & -  \\ [+0.5ex]  \hline \\[-2.3ex]  
     atmo. prompt & $<192$  & $<32$ & -\\ [+0.5ex] 
      & ($<57$)  & ($<7$) & -\\ \hline 
\end{tabular}
\caption{Number of events  for the 
six years cascade data.  The number of astrophysical neutrinos results from the single power law best fit. 
Numbers of events given in brackets refer to neutrinos with reconstructed energies above  $10\,\textnormal{TeV}$. The number of atmospheric tau neutrinos is negligible. 
Number of Glashow Resonance (astro. GR) events are evaluated assuming $pp$ type sources in the  $4-8$ PeV energy range. 
}
\label{tab4}
\end{table}
Other IceCube results are shown as blue, green and black curves for the muon neutrinos~\cite{Tracks10y},  
HESE~\cite{HESE75y} and MESE (Medium Energy Starting Events, $E\,>\,25\,{\rm{TeV}}$)~\cite{MESE2y} analyses. 
Only the muon neutrino sample is uncorrelated with cascade events from this analysis. The muon neutrino flux, measured for energies above $40\,\textnormal{TeV}$ from the Northern sky, 
is in agreement with the cascade result at the level of $1.5\sigma$  corresponding to a p-value of $0.07$.
The electron and tau neutrino (cascade) and all-neutrino flavor (HESE and MESE) measurements, which are correlated, are consistent in the overlapping energy range.

\begin{figure}[h]
\centering
\includegraphics[width=0.44\textwidth]{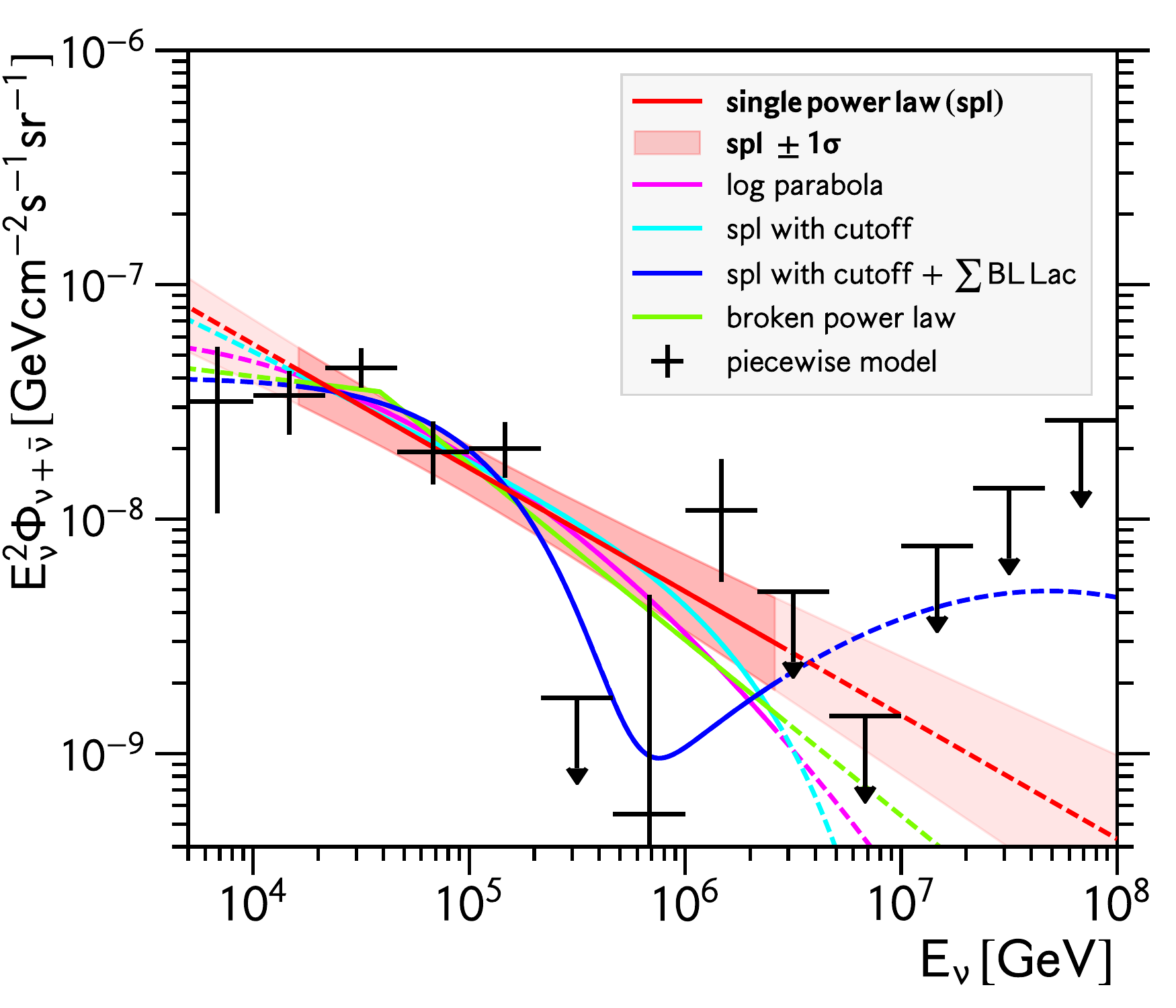}
\caption{
Astrophysical neutrino flux per neutrino flavor as a function of energy. 
Black crosses represent the differential flux model best fit results for the $2010-2015$ (6 years) cascade data.
Colored solid (dashed) curves represent astrophysical neutrino flux models in (outside of) the sensitive energy range
from $16\,\textnormal{TeV} $ to $2.6\,\textnormal{PeV}$. 
Their functional forms as well as fit results are given in Table~\ref{tab:fits}. 
The $1\sigma$ data uncertainties, data limits and uncertainty band 
correspond to the $68\%$ C.L. simultaneous coverage for the unbroken single power law flux.
}
\label{fig3}
\end{figure}

The results from fits beyond a single power-law model assumption are described below.  In the differential model  we assumed the flux 
follows an $E^{-2}$ spectrum in the individual neutrino energy segments with independent normalizations~\cite{HansThesis}.  The corresponding fit results, which indicate the strength of the astrophysical neutrino flux, are shown as black points in Fig.~\ref{fig3}. The fit results assuming other hypotheses are shown as curves with functional forms given in  
Tab.~\ref{tab:fits}. The red curve is the result of the single power law fit (hypothesis A) with the band indicating allowed parameters at 68\% C.L.  
Single power law fit results, obtained in the Southern and Northern skies separately (hypothesis F) lead to similar results.  
Other models assume additional features in the flux shape, such as a cutoff (hypotheses B and E), break in the spectrum (hypothesis D), energy dependence of the spectral index (hypothesis C) as well as an additional neutrino emission component at high neutrino energies from the population of BL Lac blazars (hypothesis E). The latter has been modeled according to~\cite{PPGR} with one free parameter, the neutrino to $\gamma$-ray intensity ratio, $Y_{\nu\gamma}$.   
 The fit results are given in Tab.~\ref{tab:fits}.   Although not statistically significant, the results (hypothesis C, D and E)  indicate an overall soft spectral index ($\gamma \sim  2.4 - 2.6$), a softening of spectral index with energy from $\gamma \sim 2.0$ to $\gamma \sim 2.75$ above $\sim40\,\textnormal{TeV}$, or a cutoff in the flux from the low energy component at energies as low as $\sim \,\textnormal{0.1}$ PeV.  The non-zero contribution from the BL Lac neutrino flux component (hypothesis E), 
which is proportional to the $Y_{\nu\gamma}$,  is statistically non-significant. 
We thus placed an upper limit on the 
ratio $Y_{\nu,\gamma} < 0.41$ at $90\%$ C.L., leading to the conclusion that a significant fraction of the $\gamma$-ray emission from BL Lacs is due to leptonic processes, in agreement with the IceCube limit at ultra high energies~\cite{ehe2018,ehe2018E}.
Current statistics are not sufficient to distinguish between models that go beyond the single power law (hypotheses B-F, Tab.~\ref{tab:fits}). 
The most significant extension to the single power law is hypothesis C, assuming energy dependent  spectral indices,  with a p-value of $0.06$.

In summary,  our results are consistent with the hypothesis that the flux of astrophysical electron and tau neutrinos follows a single power law, 
with a spectral index of $\gamma=2.53\pm0.07$ and a flux normalization for each neutrino flavor of $\phi_{astro} =(1.66^{+0.25}_{-0.27})$ at $\textnormal{E}_0=100\,\textnormal{TeV}$. 
In the measured energy range we reject spectral indices $\gamma\leq2.28$ at $\ge3\sigma$  level.
The sizable and smooth flux measured  below $\sim 100\,{\rm{TeV}}$  remains a puzzle.
In order to not violate the isotropic diffuse gamma-ray
background~\cite{fermi-lat-diffuse},  it suggests  the existence of astrophysical neutrino sources characterized by dense environments which are opaque to gamma-rays.

\vspace*{0.2cm}

\begin{acknowledgments}
The IceCube collaboration acknowledges the significant contributions to this manuscript from the Stony Brook University. 
We acknowledge the support from the following agencies:
USA {\textendash} U.S. National Science Foundation-Office of Polar Programs, 
U.S. National Science Foundation-Physics Division, 
Wisconsin Alumni Research Foundation, 
Center for High Throughput Computing (CHTC) at the University of Wisconsin-Madison, 
Open Science Grid (OSG), 
Extreme Science and Engineering Discovery Environment (XSEDE), 
U.S. Department of Energy-National Energy Research Scientific Computing Center, 
Particle astrophysics research computing center at the University of Maryland, 
Institute for Cyber-Enabled Research at Michigan State University, and Astroparticle physics computational facility at Marquette University; 
Belgium {\textendash} Funds for Scientific Research (FRS-FNRS and FWO), FWO Odysseus and Big Science programmes, 
and Belgian Federal Science Policy Office (Belspo); 
Germany {\textendash} Bundesministerium f{\"u}r Bildung und Forschung (BMBF), Deutsche Forschungsgemeinschaft (DFG), 
Helmholtz Alliance for Astroparticle Physics (HAP), 
Initiative and Networking Fund of the Helmholtz Association, 
Deutsches Elektronen Synchrotron (DESY), and High Performance Computing cluster of the RWTH Aachen; 
Sweden {\textendash} Swedish Research Council, Swedish Polar Research Secretariat, Swedish National Infrastructure for Computing (SNIC), 
and Knut and Alice Wallenberg Foundation; 
Australia {\textendash} Australian Research Council; 
Canada {\textendash} Natural Sciences and Engineering Research Council of Canada, Calcul Qu{\'e}bec, Compute Ontario, 
Canada Foundation for Innovation, WestGrid, and Compute Canada; 
Denmark {\textendash} Villum Fonden, Danish National Research Foundation (DNRF), Carlsberg Foundation; 
New Zealand {\textendash} Marsden Fund; 
Japan {\textendash} Japan Society for Promotion of Science (JSPS) and Institute for Global Prominent Research (IGPR) of Chiba University; 
Korea {\textendash} National Research Foundation of Korea (NRF); 
Switzerland {\textendash} Swiss National Science Foundation (SNSF); 
United Kingdom {\textendash} Department of Physics, University of Oxford.
\end{acknowledgments}

\bibliographystyle{apsrev}
\bibliography{prl_main}

\widetext

\newpage

\begin{table*}[h!]
\small
\centering
\begin{tabular}{ccllcc}\hline
      Hypothesis & Flux Model ($\nu_{\textnormal{astro}}$) &    $\Phi_{\textnormal{astro}}^{\nu+{\bar{\nu}}}  \left( E,\,\cos\theta\right) /C_0 =$  & Result   & g.o.f & significance $\left[ \sigma \right] $ \\  \hline \hline   \\ [-2.0ex]
	A & single power law & $\Phi_0 \left(E/E_0\right)^{-\gamma}$  & $\gamma=2.53^{+0.07}_{-0.07}$  & $0.88$ & $-$    \\[+0.5ex]
      & &  & $\Phi_0 = 1.66^{+0.25}_{-0.27}$ &   \\ [+1.5ex]
      B & single power law & $\Phi_0\left(E/E_0\right)^{-\gamma}\exp{\left(-E/E_{cut}\right)} $ &  $\gamma=2.45^{+0.09}_{-0.11}$  &$0.79$   &  $1.0$  \\[+0.5ex]
      &  with cutoff &  & $\Phi_0=1.83^{+0.37}_{-0.31}$ &    \\[+0.5ex]
      &  & & $\log_{10}(E_{cut}/\textnormal{GeV})=6.4^{+0.9}_{-0.4}$  & \\  [+1.5ex]
      C & log parabolic power law & $\Phi_0 \left(E/E_0\right)^{-\Gamma(E)}$    & $\gamma=2.58^{+0.10}_{-0.10}$   & $0.79$ & $1.6$ \\[+0.5ex]
      &  & $\Gamma\left(E\right)=\gamma+b\,\log{\left(E/E_0\right)}$  & $\Phi_0=1.81^{+0.31}_{-0.29}$ &  \\ [+0.5ex]
      &  &  & $b=0.07^{+0.05}_{-0.05}$ &  \\ [+1.5ex]
      D &   broken power law &
$\Phi_b \begin{cases}
      \left( \, E/E_b\right) ^{-\gamma_1} & E\leq E_b \\
     \left( \, E/E_b \right)^{-\gamma_2} & E>E_b   \end{cases} $
   & $\begin{aligned}&\Phi_0=1.71^{+0.65}_{-0.29}\\&\log_{10}(E_{b}/\textnormal{GeV})=4.6^{+0.5}_{-0.2}\end{aligned}$   &  $0.82$ & $1.3$ \\
   &   & 
$ \Phi_{b}=\Phi_0\times\begin{cases}
\left(E_0\,/\,E_b\right)^{\gamma_1}  & E_b > E_0 \\
\left(E_0\,/\,E_b\right)^{\gamma_2}  & E_b \le E_0 
\end{cases} $
   &   $\begin{aligned}\gamma_1=2.11^{+0.29}_{-0.67}\\\gamma_2=2.75^{+0.29}_{-0.14}\end{aligned}$&  \\[+4.0ex]
  E & single powerlaw  & $\Phi_0\left(E/E_0\right)^{-\gamma}\exp{\left(-E/E_{cut}\right)}$ & $\gamma=2.0^{+0.3}_{-0.4}$ & $0.78$ & $1.1$ \\[+0.5ex]
   &   with cutoff + $\sum$ BL Lac & $\quad + Y_{\nu\gamma}\times f\left(E\right)$  & $\Phi_0=4.3^{+3.2}_{-1.6}$ &  & \\[+0.5ex]
  &  [Padovani BLLac] & & $\begin{aligned}&\log_{10}(E_{cut}/\textnormal{GeV})=5.1^{+0.3}_{-0.2} \\ &Y_{\nu\gamma}=0.20^{+0.12}_{-0.09}\end{aligned}$   \\[+3.5ex]
 F &  two hemispheres & $\begin{cases}\Phi_N \left(E/E_0\right)^{-\gamma_N} & \cos\theta\leq 0\\ \Phi_S \left(E/E_0\right)^{-\gamma_S} & \cos\theta>0 \end{cases}$ & $\begin{aligned}&\gamma_N=2.45^{+0.17}_{-0.36}\\&\Phi_N=1.3^{+0.7}_{-1.0}\end{aligned}$ & $0.87$ & $0.0$ \\[+2.5ex]
 & &  & $\begin{aligned}&\gamma_S=2.52^{+0.10}_{-0.11}\\&\Phi_S=1.62^{+0.30}_{-0.29}\end{aligned}$  \\[+2.5ex]
  G &  single power law ($p\gamma$) &  $\Phi_0 \left(E/E_0\right)^{-\gamma}$  & $\gamma=2.50^{+0.07}_{-0.07}$  & $0.88$ & $0.7$     \\[+1.5ex]
  & &  & $\Phi_0 = 1.62^{+0.25}_{-0.27}$ &   \\[+1.5ex]
  \hline \hline \end{tabular}
\caption{
$C_0 = 3\times 10^{-18}\,\textnormal{GeV}^{-1}\cdot\textnormal{cm}^{-2}
\cdot\textnormal{s}^{-1}\cdot\textnormal{sr}^{-1}$ and $E_{0} = 100\, \textnormal{TeV}$. \newline
Fit results for different hypotheses, assuming the baseline $(\nu_e : \nu_{\mu} : \nu_{\tau})_E = (\bar{\nu}_e : \bar{\nu}_{\mu} : \bar{\nu}_{\tau} )_E  = 0.5: 0.5: 0.5$ flavor composition expected for $pp$ sources (hypotheses A-F) and  $(\nu_e : \nu_{\mu} : \nu_{\tau})_E = 0.78: 0.61: 0.61$, $(\bar{\nu}_e : \bar{\nu}_{\mu} : \bar{\nu}_{\tau} )_E = 0.22 : 0.39 : 0.39$ expected for $p\gamma$ sources (hypothesis G).
Goodness of fit (g.o.f.) test used in this work is the  saturated Poisson likelihood test~\cite{gof,HansThesis}. The corresponding g.o.f.  p-values have been calculated as described in~\cite{HansThesis} (Section $5.5$).  
Significance $\sigma$ of alternative, more complex astrophysical flux models over single power-law model as determined from toy experiments. The significance of the single-power law fit with respect to the background only hypothesis ($\Phi_{astro}=0$) is $9.9\sigma$. All significances are given using the one-sided convention.}
\label{tab:fits}
\end{table*}

\end{document}



\begin{figure*}[h!]
  \centering
    \includegraphics[width=0.43\textwidth]{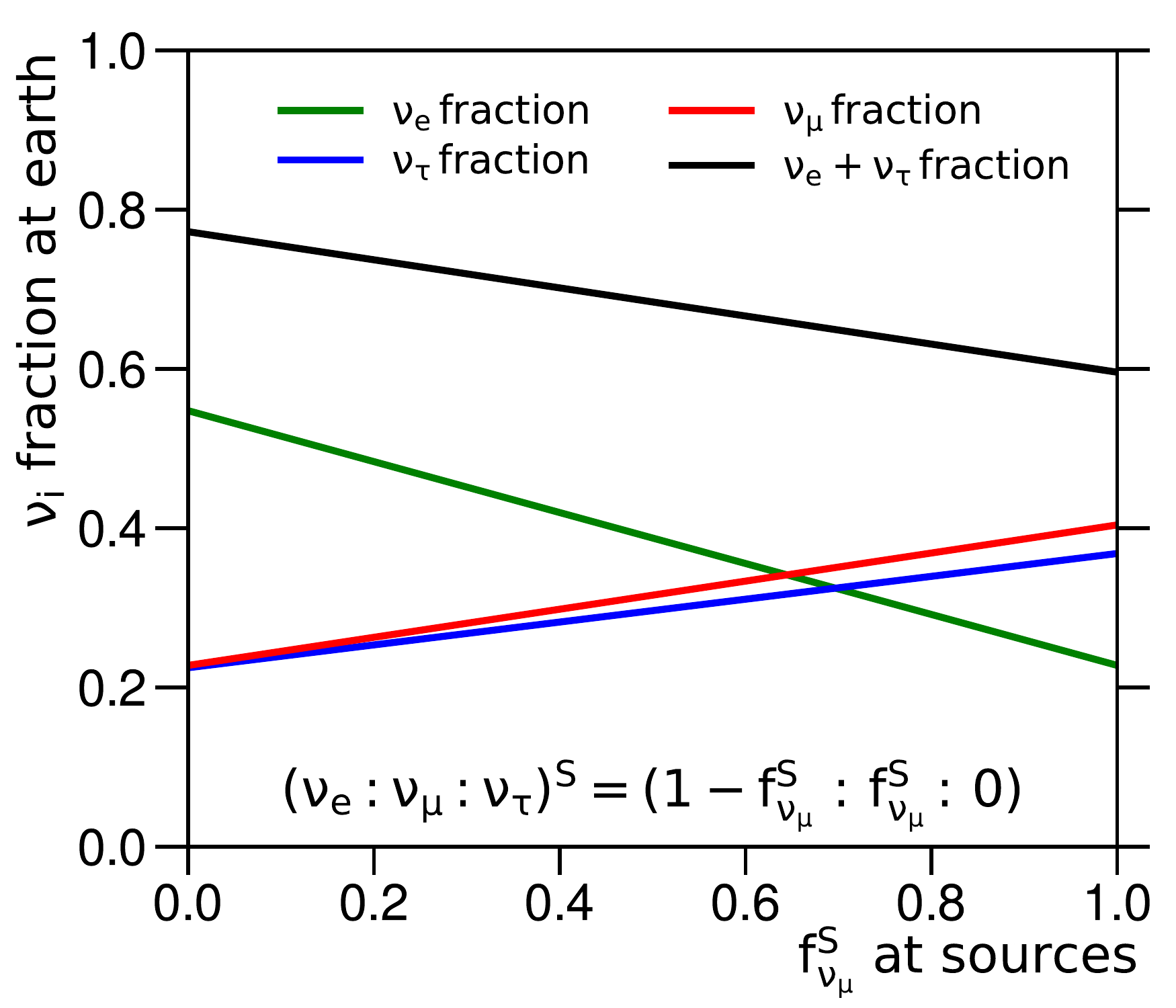}\includegraphics[width=0.43\textwidth]{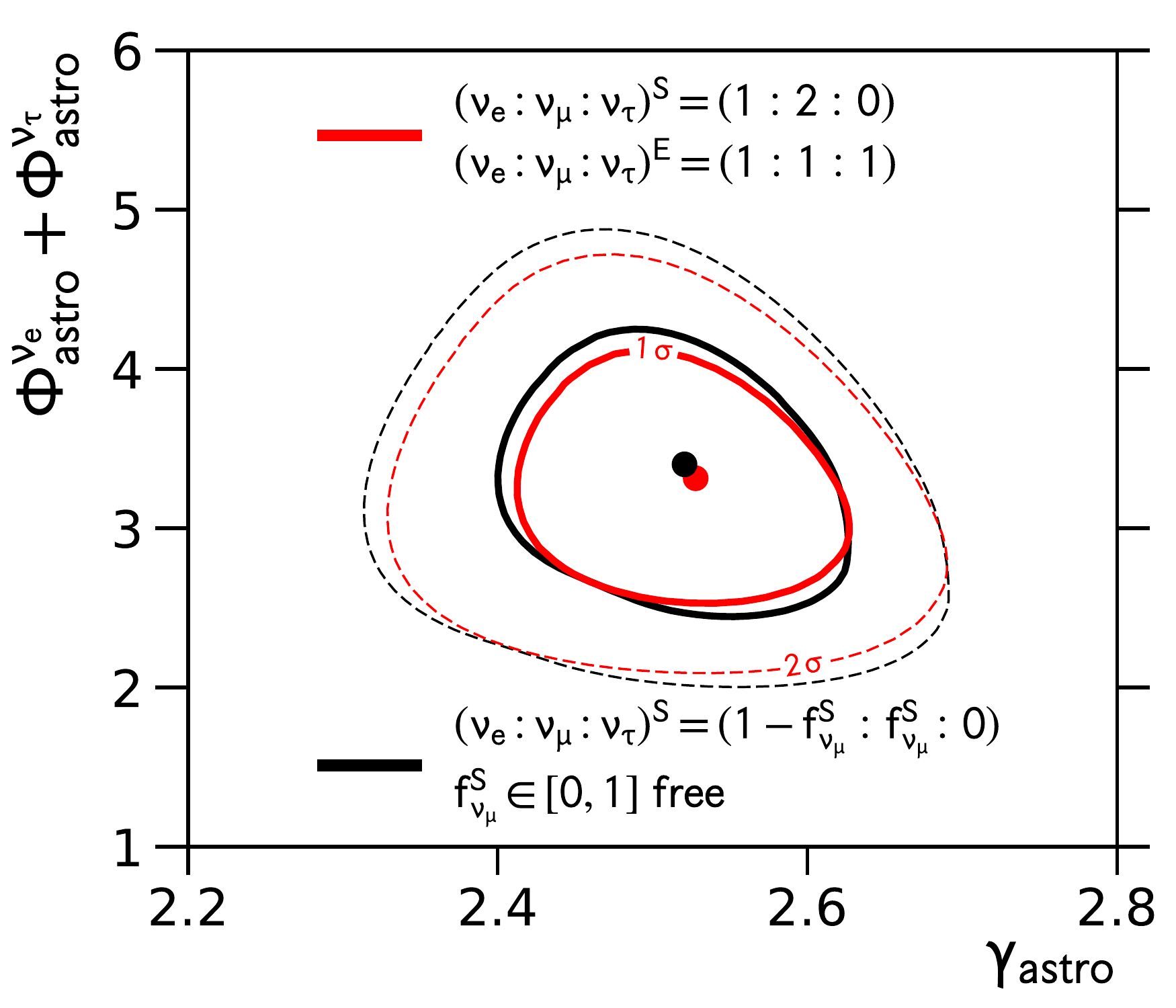}
  \caption{Left: possible contributions from each $\nu$ flavor to the total $\nu$ flux at Earth as function of the fraction of muon neutrinos $f_{\nu_\mu}^S\in[0,1]$ injected at astrophysical $\nu$ sources assuming standard neutrino oscillations 
[23,24].
Right:  measurement of the combined flux (single power-law) of electron and tau neutrinos
assuming the baseline $(\nu_e : \nu_{\mu} : \nu_{\tau})_S = 1: 2: 0$ flavor composition expected for ideal pion-decay sources (red) and 
taking into account possible variations in the injected flavor ratio at astrophysical sources through an additional nuisance parameter $f_{\nu_\mu}^S\in[0,1]$ (black).
Solid lines correspond to $68\%\,\textnormal{C.L.}$ and dashed lines correspond to $95\%\,\textnormal{C.L.}$.
Both results are approximately identical, because the contribution of astrophysical muon neutrinos to the cascade samples is suppressed.}
                \label{fig:triangle2}
\end{figure*}
